\documentclass[sigconf]{acmart}
\usepackage{multirow}
\usepackage{makecell} 
\usepackage{algorithm}
\usepackage{algpseudocode}
\usepackage{graphicx}
\usepackage{subfigure}
\usepackage{caption}
\AtBeginDocument{%
  }

\setcopyright{acmlicensed}
\copyrightyear{2018}
\acmYear{2018}
\acmDOI{XXXXXXX.XXXXXXX}
\acmConference[Conference acronym 'XX]{Make sure to enter the correct
  conference title from your rights confirmation email}{June 03--05,
  2018}{Woodstock, NY}
\acmISBN{978-1-4503-XXXX-X/2018/06}

\renewcommand\footnotetextcopyrightpermission[1]{} % removes footnote with conference information in first column
\begin{document}

\title{Zero-Shot Forecasting of Network Dynamics through Weight Flow Matching}


\author{Shihe Zhou}
\authornote{Equal Contribution}
% \email{zhoush23@mails.tsinghua.edu.cn}
\affiliation{%
  \institution{Tsinghua University}
  \city{Beijing}
  \country{China}
}

\author{Ruikun Li}
\authornotemark[1]
% \email{}
\affiliation{%
  \institution{Tsinghua University}
  \city{Beijing}
  \country{China}
}

\author{Huandong Wang}
\authornote{Corresponding authors @ wanghuandong@tsinghua.edu.cn}
% \email{wanghuandong@tsinghua.edu.cn}
\affiliation{%
  \institution{Tsinghua University}
  \city{Beijing}
  \country{China}
}

\author{Yong Li}
% \email{}
\affiliation{%
  \institution{Tsinghua University}
  \city{Beijing}
  \country{China}
}

\renewcommand{\shortauthors}{Trovato et al.}

\begin{abstract}

Forecasting state evolution of network systems, such as the spread of information on social networks, is significant for effective policy interventions and resource management. However, the underlying propagation dynamics constantly shift with new topics or events, which are modeled as changing coefficients of the underlying dynamics. Deep learning models struggle to adapt to these out-of-distribution shifts without extensive new data and retraining. To address this, we present Zero-Shot \underline{F}orecasting of \underline{N}etwork Dynamics through Weight \underline{F}low \underline{M}atching (FNFM), a generative, coefficient-conditioned framework that generates dynamic model weights for an unseen target coefficient, enabling zero-shot forecasting. Our framework utilizes a Variational Encoder to summarize the forecaster weights trained in observed environments into compact latent tokens. A Conditional Flow Matching (CFM) module then learns a continuous transport from a simple Gaussian distribution to the empirical distribution of these weights, conditioned on the dynamical coefficients. This process is instantaneous at test time and requires no gradient-based optimization. Across varied dynamical coefficients, empirical results indicate that FNFM yields more reliable zero-shot accuracy than baseline methods, particularly under pronounced coefficient shift.

\end{abstract}






\maketitle

\section{Introduction}
The propagation of behaviors, evolution of cultural norms, and even the formation of consensus on social networks can all be modeled as spatiotemporal dynamic processes among individuals on complex networks~\cite{jusup2022social, zheng2013spreading, li2024predicting}. 
Accurately forecasting the evolution of these dynamics is crucial for understanding sociophysical phenomena and for key applications such as curbing the spread of misinformation~\cite{bovet2019influence, iacopini2024temporal, meng2025spreading}.

\begin{figure*}[!ht]
    \centering
    \includegraphics[width=\textwidth]{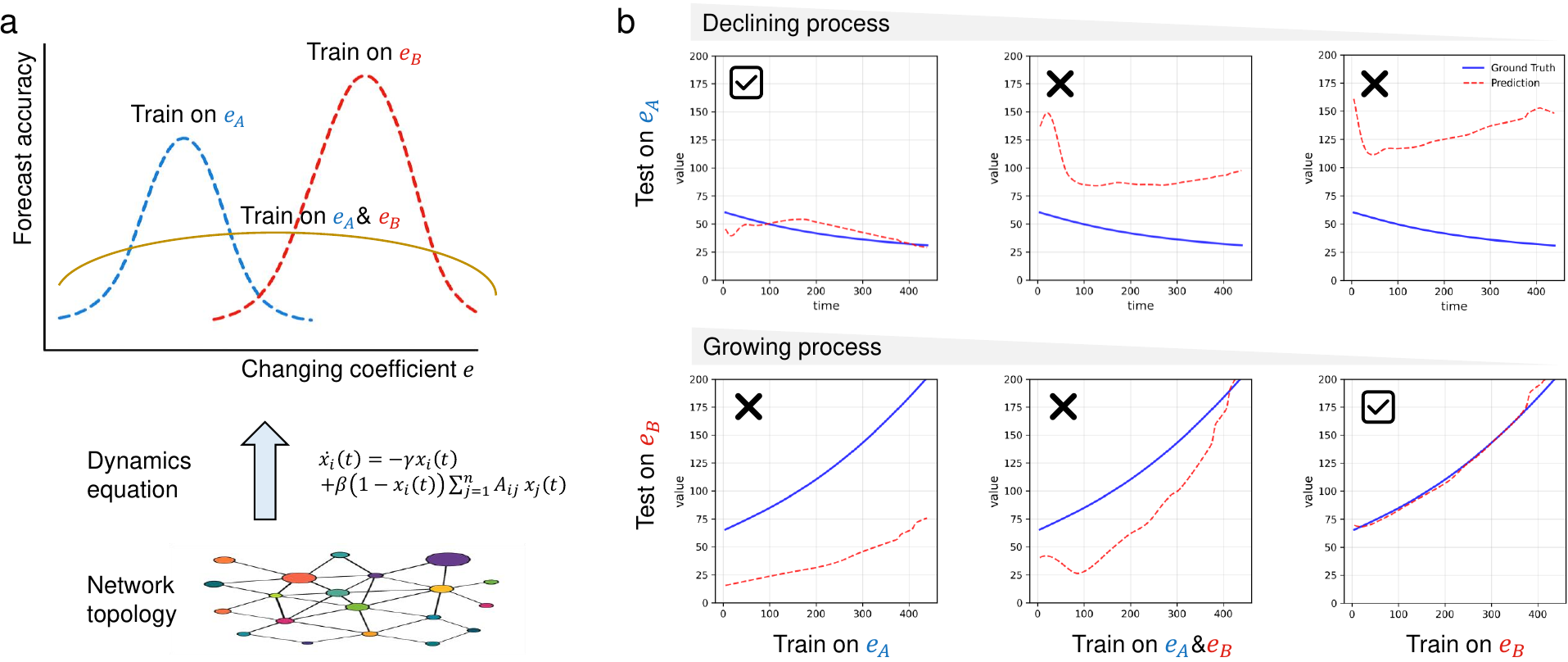} 
    \vspace{-0.7cm}
    \caption{
    The generalization trap in network dynamics.
    (a) A generalist model trained on mixed data struggles to outperform specialized expert models. 
    Network dynamics adopts the classic information dissemination model~\cite{vespignani2012modelling}, in which the dynamic behavior is governed by the popularity coefficient.
    (b) Training and testing performance on cross-environment propagation dynamics, where $e_A$ and $e_B$ are propagation processes with different coefficients.
    % prediction results on environment $e_A$(top row, $\beta$=0.02, $\gamma$=0.016) and $e_B$ (bottom row, $\beta$=0.02, $\gamma$=0.068); columns 1–3 are models trained on $e_A$, $e_A \& e_B$, and $e_B$.
    }
    \label{fig:dynamic_variety}
\end{figure*}

The complexity of network dynamics stems from the intricate interplay between network topology and the parameters of the underlying dynamics. 
Even with identical topologies and governing equations, subtle shifts in dynamic parameters can push a system toward entirely different critical regimes, fundamentally altering its propagation behavior~\cite{vespignani2012modelling,gomez2018critical,hensspatiotemporal2019, liu2023emergence}. 
As we demonstrate on a social media information propagation model (Figure~\ref{fig:dynamic_variety}a), a mere difference in popularity coefficients (define as $\frac{\beta}{\gamma}$) leads two propagation trajectories toward starkly different outcomes: rapid decay versus viral spread. 
This phenomenon precisely captures the disparity in how opinions on different topics propagate through cyberspace, while also posing a stringent challenge to the generalization capability of predictive models: they must be able to accurately forecast dynamic evolution in new environments.

To train generalizable models from a limited set of observed environments, existing work predominantly follows two paths. 
The first path involves building "one-for-all" spatio-temporal foundation models, attempting to train a universal predictor by aggregating data from all environments~\cite{goodge2025spatio,liang2025foundation,yuan2024urbandit,li2024opencity}. 
However, these monolithic models, guided by the principle of empirical risk minimization, often achieve generalization at the expense of specialized performance, leading to performance on specific tasks that can be inferior to that of much smaller expert models (as shown in Figure~\ref{fig:dynamic_variety}b). 
The second path is based on meta-learning approaches, which rapidly adapt a model to new environments, thereby reducing data dependency~\cite{pan2020spatio,xu2025spatio,pan2019urban,qin2021robust}. 
However, meta-learning frameworks still rely on the availability of at least a small amount of historical trajectory data from the target environment for finetuning. 
This precondition, however, does not hold in many high-value predictive scenarios~\cite{ferguson2020report,lazer2009computational,halloran2008modeling,forrester1970urban}. In such scenarios, a decision-maker might need to predict the potential consequences of a hypothetical environmental coefficient (e.g., the adoption rate of a new policy). 
Therefore, how to perform reliable zero-shot prediction for network dynamics under new environmental coefficients remains a critical open question.

In this paper, we introduce \underline{F}orecasting of \underline{N}etwork Dynamics through Weight \underline{F}low \underline{M}atching (FNFM), a novel generative framework that addresses the challenge of zero-shot prediction for network dynamics across varying environments. 
Instead of predicting trajectories directly, FNFM learns to generate the complete \textit{weights} of a specialized forecaster model tailored to any given environmental coefficients. 
FNFM first collects a diverse set of expert weights from various seen environments. It then employs a Variational Autoencoder (VAE) to learn a compact and smooth latent manifold of these weights. Finally, a Conditional Flow Matching (CFM) model is trained to map environmental coefficients to this manifold, enabling the conditional synthesis of new latent vectors. 
At inference time, this process is instantaneous and requires no finetuning, making FNFM a powerful tool for forecasting for novel scenarios on demand.

Our main contributions are summarized as follows:
\begin{itemize}
    \item We propose a new paradigm for zero-shot forecasting of network dynamics, shifting the objective from trajectory prediction to the direct generation of model weights.
    \item We introduce FNFM, a novel framework that operationalizes this paradigm by synergistically combining a VAE and a Conditional Flow Matching model to learn the complex mapping from dynamic coefficients to optimal model weights.
    \item We conduct extensive experiments demonstrating that FNFM significantly outperforms state-of-the-art baselines by an average of 8.30\% in zero-shot forecasting scenarios, showcasing its superior generalization.
\end{itemize}

\section{Preliminary}
\subsection{Problem Definition}
We consider a dynamic process evolving over a network of $n$ nodes, where each node possesses a $d$-dimensional feature vector.
A core challenge in forecasting such dynamics is that while different environments may share the same underlying network topology and governing equations, they are distinguished by a set of dynamic coefficients.
These coefficients, denoted by an environmental vector $e \in E$, critically alter the system's behavior, leading to fundamentally different temporal evolution patterns.

Formally, given an adjacency matrix $A \in \mathbb{R}^{n \times n}$ and the environmental coefficient vector $e$, the network dynamics can be described by a system of ordinary differential equations (ODEs):
$$
\frac{d\mathbf{X}(t)}{dt} = F(\mathbf{X}(t), A, e)
$$
where $\mathbf{X}(t) = (x_1(t), \dots, x_n(t))^T$ represents the state of all nodes at time $t$, and the nonlinear function $F$ is parameterized by the environment $e$.

Our task is zero-shot forecasting. We assume access to a set of historical trajectories collected from a number of \textit{seen} environments, $E_{seen} \subset E$.
The objective is to train a model that can accurately predict the future trajectory for a previously \textit{unseen} environment $e_{unseen} \in E_{unseen}$, where the seen and unseen environment sets are disjoint ($E_{seen} \cap E_{unseen} = \emptyset$).
Specifically, for a given trajectory, the forecasting task is defined as predicting a future window of states $\mathbf{X}_{t+1:t+N}$ given an observed historical window $\mathbf{X}_{t-H+1:t}$, where $H$ is the look-back window size and $N$ is the prediction horizon.

\subsection{Conditional Flow Matching}
Flow Matching is a powerful and recently developed generative modeling framework designed to learn a transformation from a simple prior distribution, $p_0$, to a complex data distribution, $p_1$ \cite{lipmanflow2023,lipman2022flow, ma2024sit}. This is achieved by training a parameterized, time-dependent vector field, $v_\xi(x, t)$, that learns to match a target velocity field guiding the transformation. Conditional Flow Matching (CFM) extends this concept by allowing the transformation to be dependent on a conditioning variable, $c$. The goal is thus to learn a map from $p_0$ to a conditional target distribution $p_1(x|c)$.

While various path definitions are possible, a common and effective approach is to use a straight-line path between samples from the source and target distributions \cite{tongimproving2024}. Specifically, for a pair of samples $x_0 \sim p_0$ and $x_1 \sim p_1( \cdot | c)$, the probability path $p_t(x|x_0, x_1)$ is defined as a Gaussian bridge:
\begin{equation}
    p_t(x|x_0, x_1) = \mathcal{N}(x \mid (1-t)x_0 + t x_1, \sigma^2),
    \label{eq:p_t_prelim}
\end{equation}
where $t \in [0, 1]$ and $\sigma^2$ is a small variance. A key advantage of this formulation is that the corresponding target velocity field simplifies to a constant vector:
\begin{equation}
    u_t(x|x_0, x_1) = x_1 - x_0.
    \label{eq:u_t_prelim}
\end{equation}
This provides a direct and stable regression target for the conditional neural network $v_\xi(x, t, c)$. The network's weights $\xi$ are optimized by minimizing the following loss function:
\begin{equation}
    \mathcal{L}_{CFM}(\xi) = \mathbb{E}_{t, c, x_0, x_1} \left[ \left\| v_\xi\left((1-t)x_0 + t x_1, t, c\right) - (x_1 - x_0) \right\|^2 \right],
\end{equation}
where the expectation is taken over time $t \sim \mathcal{U}(0,1)$, the conditioning variable $c$, prior samples $x_0 \sim p_0$, and target samples $x_1 \sim p_1(x|c)$. To further improve efficiency, modern implementations often pair samples $x_0$ and $x_1$ using mini-batch optimal transport (OT) plans, resulting in shorter and more direct flows \cite{tongimproving2024}.

\begin{figure*}[!ht]
    \centering
    \includegraphics[width=0.92\textwidth]{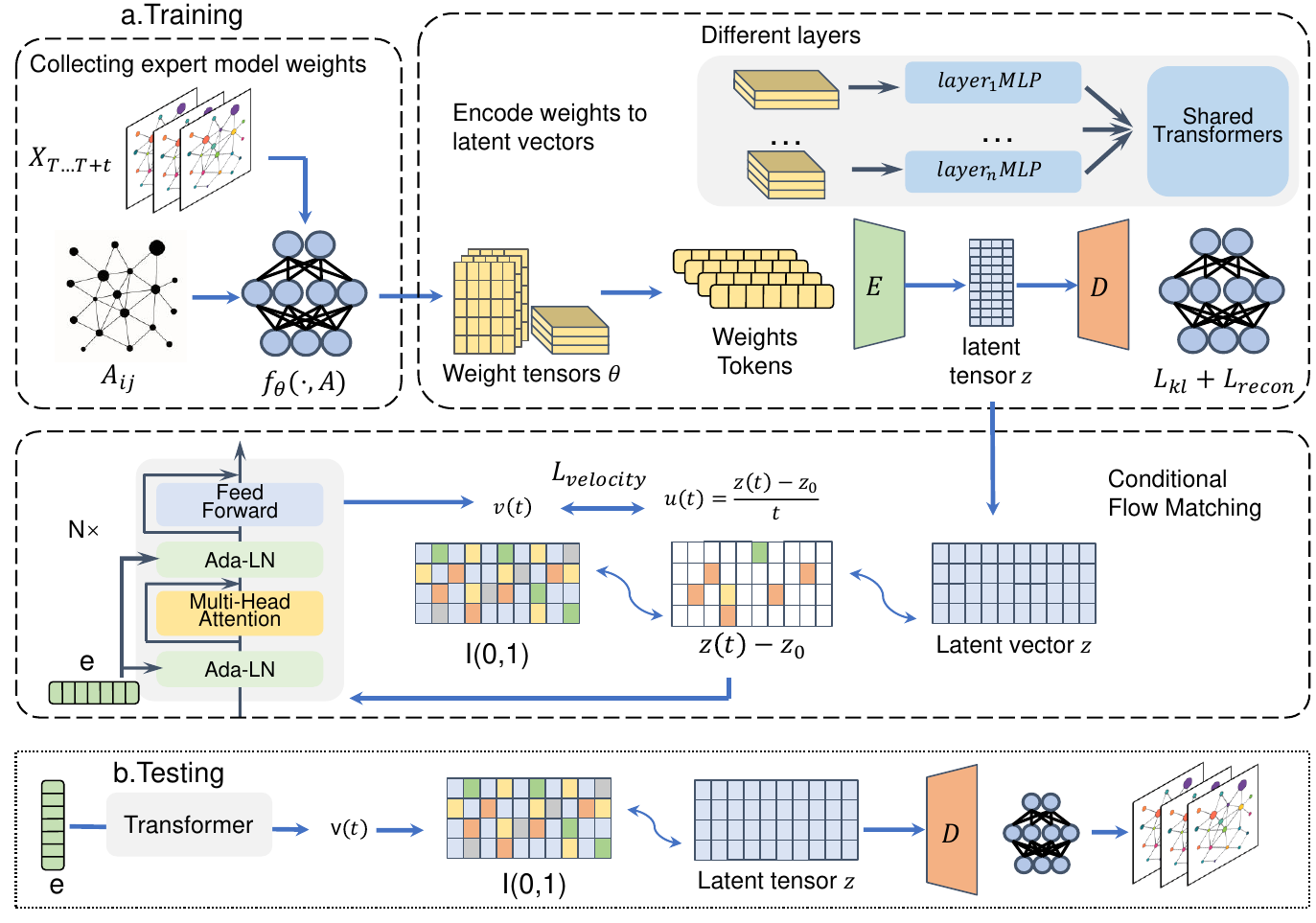}
    \vspace{-0.3cm}
    \caption{Overview of the Model Architecture. The framework comprises collecting expert model weights, tokenizing and encoding the model weight to latent vectors and conditional flow matching, working synergistically for zero-shot weight generation and dynamics forecasting.}
    \label{fig:model_architecture}
\end{figure*}

\section{Methodology}
To address the challenge of zero-shot forecasting, we introduce \underline{F}orecasting of \underline{N}etwork Dynamics through Weight \underline{F}low \underline{M}atching (FNFM), a novel generative framework. 
The core paradigm of FNFM shifts from directly predicting dynamic trajectories to generating the \textit{weights} of a specialized forecaster model tailored to any given environmental condition.
As illustrated in Figure~\ref{fig:model_architecture}, our methodology accomplishes this through a three-fold pipeline. 
FNFM first collects weights from expert models, then uses a variational autoencoder to map them into a latent space, and finally trains a conditional flow matching model to generate latent vectors for the zero-shot synthesis of new models.

\subsection{Collecting Expert Model Weights}
We conceptualize the optimized weights of an expert forecaster trained on a single environment as a high-dimensional vector that captures the essence of that environment's unique network dynamics. 
Our FNFM framework is designed to learn the joint distribution of these weights and their corresponding environmental coefficients in a data-driven manner.
Therefore, the foundational step of our methodology is to construct a dataset of these expert weights for all \textit{seen} environments.

To achieve this, for each seen environment $e \in E_{seen}$ and its associated trajectory data $X^{(e)}$, we train a dedicated forecaster to parameterize the dynamical function $F_\theta(\cdot, A)$. 
While our framework is agnostic to the specific forecaster architecture, we employ a Spatio-Temporal Graph Convolutional Network (STGCN) \cite{yuspatiotemporal2018} in our implementation. 
The weights $\theta^{(e)}$ for each expert are optimized by minimizing a multi-step forecasting loss over sliding windows of length $H$ with a prediction horizon of $N$:
\begin{equation}
    \theta^{(e)} = \arg\min_{\theta} \sum_{t=H}^{T-H-N} \left\| F_{\theta}(X^{(e)}_{t-H+1:t}, A) - X^{(e)}_{t+1:t+N} \right\|^2,
    \label{eq:expert_loss}
\end{equation}
where $\theta^{(e)}$ denotes the resulting weights of the expert model tailored to environment $e$.

Each expert model is trained to converge on its specific environmental data using the Adam optimizer \cite{kingma2014adam}. The final collection of optimized weights, $\{ \theta^{(e)} \mid e \in E_{seen} \}$, serves as the target data for the subsequent generative learning stages of our framework.

\subsection{Weight Sequence Tokenizer}
We treat the weights of a neural network as a novel data modality. Fundamentally, these weights constitute a complex, structured representation, not merely a flat vector of numbers. To make this data compatible with powerful sequence-based models (like Transformers) while preserving the network's inherent architectural inductive biases, we introduce a tokenization scheme guided by the data flow through the network's computational units.

Our process operates on the fundamental building blocks of most neural networks: convolutional and linear layers~\cite{kofinasgraph2024, li2025predicting, li2025weightflow}. For each expert model's weights, we decompose them layer by layer into a sequence of meaningful tokens.

\noindent\textbf{Convolutional Layers.} For a convolutional layer \(\ell\) with kernel tensor \(\Omega_\ell \in \mathbb{R}^{C_{\text{out}} \times C_{\text{in}}\times h\times w}\), we form one token for each output channel. Each token aggregates all the weights responsible for producing that single output channel's feature map. Concretely, the token for the \(o\)-th output channel is:
\[
\mathbf{w}^{(\ell)}_{o} = \operatorname{flatten}\!\big(\Omega_{\ell,o,:,:,:}\big)\in\mathbb{R}^{C_{\text{in}} \cdot h \cdot w},\quad \text{for } o=1,\ldots,C_{\text{out}}.
\]

\noindent\textbf{Linear Layers.} Similarly, for a linear layer \(\ell\) with weight matrix \(\mathbf{W}_\ell\in\mathbb{R}^{D_{\text{out}}\times D_{\text{in}}}\) and bias vector $\mathbf{b}_\ell\in \mathbb{R}^{D_{\text{out}}}$, we define one token per output unit (neuron). This token includes all incoming weights and the bias for that unit:
\[
\mathbf{w}^{(\ell)}_{o} = [\mathbf{W}_{\ell,o,:}; \mathbf{b}_{\ell,o}]\in\mathbb{R}^{D_{\text{in}}+1},\quad \text{for } o=1,\ldots,D_{\text{out}}.
\]
This procedure losslessly transforms the entire weights $\theta$ of a $L$-layer neural network into an ordered sequence of tokens: 
\[
\{ \mathbf{w}^{(1)}_1, \dots, \mathbf{w}^{(1)}_{C_{out}}, \dots, \mathbf{w}^{(L)}_1, \dots, \mathbf{w}^{(L)}_{D_{out}} \}. 
\]
Each token represents a self-contained computational unit, and the sequence preserves the layer-wise structure of the original model. This tokenized sequence serves as the direct input for our subsequent generative modeling stage.

\subsection{Weight Variational Autoencoder}
To facilitate stable and effective generative learning, we first compress the high-dimensional weight token sequence into a smooth and compact low-dimensional latent space~\cite{dao2023flow}. We achieve this using a purpose-built Variational Autoencoder (VAE) featuring a Transformer-based architecture.

\subsubsection{Model Architecture}
The VAE consists of an encoder $E$ that maps a sequence of weight tokens to a latent vector $z$, and a decoder $D$ that reconstructs the token sequence from that vector.

\noindent\textbf{Layer-wise Token Embedding.} The raw tokens $\{ \mathbf{w}^{(\ell)}_{o} \}$ from different layers possess varying dimensionalities, which is incompatible with a standard Transformer. To handle this, we first employ a set of layer-wise projection networks (MLPs), $f_\ell$, to map each raw token into a fixed-dimensional embedding space:
\begin{equation}
    \mathbf{h}^{(\ell)}_{o} = f_\ell(\mathbf{w}^{(\ell)}_{o}) \in \mathbb{R}^{d_{model}}.
\end{equation}

\noindent\textbf{Transformer Encoder.} The resulting uniform-sized embeddings are fed into a multi-block Transformer encoder. Each block applies multi-head self-attention followed by a position-wise feed-forward network with residual connections and layer normalization:
\begin{align}
    \mathbf{A} &= \text{Concat}(\text{head}_1, \dots, \text{head}_k) \mathbf{W}^O, \\
    \text{where } \text{head}_i &= \text{Attention}(\mathbf{H}\mathbf{W}_i^Q, \mathbf{H}\mathbf{W}_i^K, \mathbf{H}\mathbf{W}_i^V), \\
    \mathbf{H}' &= \text{LayerNorm}(\mathbf{H} + \mathbf{A}), \\
    \mathbf{H}_{out} &= \text{LayerNorm}(\mathbf{H}' + \text{FFN}(\mathbf{H}')).
\end{align}
The final representation of each token is then passed through two separate linear layers to parameterize the mean $\mu$ and log-variance $\log\sigma^2$ of the approximate posterior distribution $q_{\phi}(\mathbf{z}|\mathbf{w})$. A latent vector $\mathbf{z}$ is then sampled using the reparameterization trick.

\noindent\textbf{Transformer Decoder.} The decoder mirrors the encoder's architecture. It takes the latent vector $\mathbf{z}$ as a global conditioning input and reconstructs the sequence of embeddings. Finally, a set of layer-wise output networks, $g_\ell$, project the decoder's output embeddings from $\mathbb{R}^{d_{model}}$ back to their original, layer-specific token dimensions to produce the reconstructed weights $\hat{\mathbf{w}}$.

\subsubsection{Training Objective}
Let $\phi$ and $\psi$ represent the learnable parameters of encoder $E$ and decoder $D$.
The entire VAE is trained end-to-end by maximizing the Evidence Lower Bound (ELBO) on the log-likelihood of the weights:
\begin{equation}
    \mathcal{L}_{\text{ELBO}}(\mathbf{w};\phi,\psi) = \mathbb{E}_{q_{\phi}(\mathbf{z}|\mathbf{w})} \left[ \log p_{\psi}(\mathbf{w}|\mathbf{z}) \right] - \beta \cdot \text{KL} \left[ q_{\phi}(\mathbf{z}|\mathbf{w}) \,\|\, p(\mathbf{z}) \right].
\end{equation}
The objective consists of two key terms. The first is the reconstruction loss, which measures the fidelity between the original and reconstructed weights, implemented as the negative mean squared error. The second is the Kullback-Leibler (KL) regularizer, which encourages the learned latent distribution $q_{\phi}(\mathbf{z}|\mathbf{w})$ to align with a simple prior $p(\mathbf{z})$, typically a standard Gaussian $\mathcal{N}(0,I)$. This regularization ensures that the latent space is smooth and well-structured, which is crucial for the subsequent generative process.

Upon convergence, the trained encoder $E_\phi$ provides a robust mapping from any set of high-dimensional expert weights $\mathbf{w}$ to a compact latent representation $\mathbf{z}$. This collection of latent vectors, $\{ \mathbf{z}^{(e)} \mid e \in E_{seen} \}$, forms the target data manifold for our conditional flow matching module.

\subsection{Conditional Flow Matching}
With the VAE encoder providing a mapping to a structured latent space, the final stage of our framework is to learn a conditional generative model within this space. The goal is to synthesize a novel latent vector $\mathbf{z}_{new}$ that corresponds to a previously unseen environmental coefficient $e_{new}$. We achieve this by training and deploying a conditional vector field using the flow matching principles outlined in the preliminaries.

\subsubsection{Training the Conditional Vector Field}
We train a time-dependent conditional vector field, parameterized by a neural network $v_\xi(\mathbf{z}, t, e)$, to approximate the target velocity field $(\mathbf{z}^{(e)} - \mathbf{z}_0)$ defined in Equation~\ref{eq:u_t_prelim}. The network's parameters $\xi$ are optimized by minimizing the following objective:
\begin{equation}
    \mathcal{L}_{CFM}(\xi) = \mathbb{E}_{t, e, \mathbf{z}_0, \mathbf{z}^{(e)}} \left[ \left\| v_\xi\left((1-t)\mathbf{z}_0 + t \mathbf{z}^{(e)}, t, e\right) - (\mathbf{z}^{(e)} - \mathbf{z}_0) \right\|^2 \right],
    \label{eq:cfm_loss}
\end{equation}
where the expectation is over time $t \sim \mathcal{U}(0,1)$, seen environments $e \sim E_{seen}$, prior samples $\mathbf{z}_0 \sim \mathcal{N}(0, I)$, and their corresponding target latent codes $\mathbf{z}^{(e)} = E_\phi(\mathbf{w}^{(e)})$.

The vector field $v_\xi$ is implemented using a Transformer architecture. To inject the environmental information $e$ effectively, we employ an Adaptive Layer Normalization (AdaLN) mechanism. Within each Transformer block, the input sequence $H_n$ is modulated before the self-attention layer:
\begin{equation}
    \text{AdaLN}(H_n, e) = \gamma(e) \odot \text{LayerNorm}(H_n) + \beta(e),
\end{equation}
where the scale $\gamma(e)$ and shift $\beta(e)$ are vectors produced from the environmental coefficient $e$ by small multi-layer perceptrons. This allows the network's behavior to be dynamically controlled by the target environment.

\subsubsection{Zero-Shot Weight Generation via Inference}
At inference time, FNFM generates a specialized set of weights for any unseen environment $e_{new}$ in a zero-shot fashion. This generation process is framed as solving an ordinary differential equation (ODE) initial value problem. Starting with a random sample $\mathbf{z}_0 \sim \mathcal{N}(0, I)$, we integrate the learned vector field $v_\xi$ from $t=0$ to $t=1$:
\begin{equation}
    \frac{d\mathbf{z}_t}{dt} = v_\xi(\mathbf{z}_t, t, e_{new}), \quad \text{with initial value } \mathbf{z}_0.
\end{equation}
This ODE is solved numerically using a standard solver such as forward Euler. For $N$ integration steps, the update rule is:
\begin{equation}
    \mathbf{z}_{k+1} = \mathbf{z}_k + \frac{1}{N} v_\xi(\mathbf{z}_k, \frac{k}{N}, e_{new}), \quad \text{for } k=0,\dots,N-1.
\end{equation}
The resulting vector at the final step, $\mathbf{z}_N \approx \mathbf{z}_1$, is the synthesized latent representation for the new environment. This vector is then passed through the pre-trained VAE decoder $D$ to generate the final, ready-to-use forecaster weights $\hat{\mathbf{w}}_{new} = D(\mathbf{z}_N) $.
The entire process is described in Algorithm~\ref{alg:cfm}.

\begin{algorithm}[!ht]
\caption{CFM for Conditional Latent Vector Generation}
\label{alg:cfm}
\begin{algorithmic}
\Require{Pre-computed latent vectors and conditions $\mathcal{Z} = \{ (\mathbf{z}^{(e)}, e) \}_{e \in E_{seen}}$}
\Ensure{Trained CFM network $v_\xi$}

\Statex \textcolor{gray}{\textit{// Stage 1: Training the Conditional Vector Field}}
\Procedure{TrainCFM}{$\mathcal{Z}$}
    \State Initialize CFM network parameters $\xi$.
    \For{each training step}
        \State Sample batch $\{(\mathbf{z}_1^{(i)}, e^{(i)})\}_{i=1}^B$ from $\mathcal{Z}$.
        \State Sample priors $\mathbf{z}_0^{(i)} \sim \mathcal{N}(0, I)$ and times $t^{(i)} \sim \mathcal{U}(0,1)$.
        \State $\mathbf{z}_t^{(i)} \leftarrow (1-t^{(i)})\mathbf{z}_0^{(i)} + t^{(i)} \mathbf{z}_1^{(i)}$ \Comment{Construct path points}
        \State $\mathbf{u}^{(i)} \leftarrow \mathbf{z}_1^{(i)} - \mathbf{z}_0^{(i)}$ \Comment{Define target velocities}
        \State $\mathcal{L}_{CFM} \leftarrow \frac{1}{B}\sum_{i=1}^B \| v_\xi(\mathbf{z}_t^{(i)}, t^{(i)}, e^{(i)}) - \mathbf{u}^{(i)} \|^2$ \Comment{Loss}
        \State Update $\xi$ by descending the gradient of $\mathcal{L}_{CFM}$.
    \EndFor
\EndProcedure

\Statex
\Statex \textcolor{gray}{\textit{// Stage 2: Generating Latent Vectors via Inference}}
\Procedure{GenerateLatentVector}{$v_\xi, e^{\text{new}}, N$}
    \State Sample prior $\mathbf{z}_0 \sim \mathcal{N}(0, I)$.
    \For{$k=0, \dots, N-1$}
        \State $\mathbf{z}_{k+1} \leftarrow \mathbf{z}_k + \frac{1}{N} \cdot v_{\xi}(\mathbf{z}_k, k/N, e^{\text{new}})$ \Comment{Forward Euler step}
    \EndFor
    \State \textbf{return} $\mathbf{z}_N$ \Comment{Latent vector for new environment}
\EndProcedure
\end{algorithmic}
\end{algorithm}

\begin{table*}[t]
    \caption{Average RMSE ($\pm$ std from 5 runs) in various environments (split shown in the first row). Best in bold, underlined for suboptimal. STGFSL adopts a parameter-free meta learning strategy, so it has no additional parameters}
    \label{tab:average_rmse}
    \centering
    \vspace{-0.3cm}
    \setlength{\tabcolsep}{3pt}
    \resizebox{\textwidth}{!}{%
    \begin{tabular}{lc*{4}{cc}}
        \toprule
        \multirow{2}{*}{Methods} & \multirow{2}{*}{Params}
        & \multicolumn{2}{c}{Hill}
        & \multicolumn{2}{c}{Epidemic}
        & \multicolumn{2}{c}{Twitter}
        & \multicolumn{2}{c}{Collab} \\
        & & In-domain & Out-domain & In-domain & Out-domain & In-domain & Out-domain & In-domain & Out-domain \\
        \midrule
        STGCN\cite{yuspatiotemporal2018} & 13M 
            & \underline{14.4060$\pm$2.3449} &  \textbf{8.2359$\pm$0.8919}
            &  0.3797$\pm$ 0.0166 & 0.2437$\pm$0.0189
            & 0.4402 $\pm$0.0400 &0.3156$\pm$ 0.0112 
            & 0.9153 $\pm$0.0613 & 0.8945$\pm$0.0618 \\
        STEP \cite{shaopretraining2022}& 11M 
            & 15.4436 $\pm$0.6322 & 13.6226$\pm$0.5231
            & \underline{0.0639$\pm$0.0021} &  \underline{0.0609$\pm$0.0042}
            & \underline{0.0612$\pm$ 0.0014} & \underline{0.0593 $\pm$0.0013}
            & \underline{0.0556$\pm$0.0033} & 0.0303$\pm$0.0041 \\
        % \midrule
        STGFSL\cite{luspatiotemporal2022} & -
            & 53.2770 $\pm$5.5370 & 38.7490 $\pm$5.3160
            & 0.1875$\pm$0.0005 & 0.3710 $\pm$0.0008
            &0.3040 $\pm$0.0010 & 0.4820  $\pm$0.0010
            & 0.0660 $\pm$0.0007 & 0.0330 $\pm$0.0012 \\
        GPD \cite{yuanspatiotemporal2024} & 12M 
            & 98.1748$\pm$0.0963 & 9.92475$\pm$1.315
            & 0.1888  $\pm$ 0.0002 & 0.0708   $\pm$ 0.0021
            & 0.1336  $\pm$0.0004 & 0.0676  $\pm$0.0020
            & 0.0722 $\pm$ 0.0001 & \underline{0.0280 $\pm$0.0004}\\
        Ours & 11M
            & \textbf{13.8942 $\pm$2.4240} & \underline{8.5595 $\pm$ 0.4594}
            & \textbf{0.0562 $\pm$0.0071}  & \textbf{0.0561$\pm$0.0059}
            & \textbf{0.0579 $\pm$ 0.0081} & \textbf{0.0512$\pm$0.0048}
            & \textbf{0.0475  $\pm$ 0.0036} & \textbf{0.0244$\pm$0.0021}\\
        \multicolumn{2}{l}{Percentage} & 3.55\% & -3.93\% & 12.05\% & 7.89\% & 5.39\% & 13.66\% & 14.57\% & 12.86\%  \\
        \midrule
        \multicolumn{2}{l}{One-per-Env STGCN}
            & 11.2409 & 8.5127 
            & 0.0496 &0.0538
            & 0.0364  & 0.0326
            & 0.0461& 0.0229\\
        \bottomrule
    \end{tabular}
    }
\end{table*}

\section{Experiment}

\subsection{Evaluation Protocol}
We evaluate our method on the task of forecasting networked dynamical systems under distribution shifts. We regard an \textbf{environment} as the combination of a specific network's trajectory data and its associated dynamic coefficients. 
For each dataset, we partition the available environments into training, validation, and testing sets, ensuring no overlap. The core of our evaluation lies in the test set, which is further divided into two distinct regions to rigorously assess generalization:
\begin{itemize}
    \item \textbf{In-Domain:} Test environments whose dynamic coefficients are \textit{interpolated} from within the range of coefficients observed during training.
    \item \textbf{Out-of-Domain:} Test environments whose dynamic coefficients are \textit{extrapolated} beyond the range of the training set coefficients. This presents a more challenging test of a model's generalization capabilities.
\end{itemize}

\subsection{Implementation Details}
We report multi-step forecasting performance using Root Mean Squared Error (RMSE) computed on the non-standardized trajectories. Across all experiments, we set the historical look-back window to $H = 50$ and the prediction horizon to $N = 50$. All model hyperparameters, for both our method and the baselines, are tuned on a dedicated validation set of environments. To ensure robust results, all reported metrics are the average of 5 independent runs using different random seeds but identical environment splits.

\subsection{Datasets}
We evaluate FNFM on four datasets designed to span both synthetic and real-world topologies, as well as heterogeneous dynamics. Each environment within a dataset is characterized by a unique dynamic coefficient.
\begin{itemize}
    \item \textbf{Hill:} Barabási–Albert (BA) synthetic networks coupled with Hill-type dynamics. This dataset contains 400 distinct environments generated by varying both topology and dynamical coefficients.
    \item \textbf{Epidemic:} The real European road network topology with SIS epidemic dynamics simulated on the graph. 121 environments are formed by varying initial conditions and transmission/recovery coefficients.
    \item \textbf{Twitter:} The real Twitter network with information propagation modeled by an SIS-like process. 121 environments are created by varying propagation rates.
    \item \textbf{Collab:} A network of scientific collaborations among institutions in New Zealand modeled by an SIS-like process. 40 environments are created by perturbing propagation coefficients.
\end{itemize}
Across all datasets, the training, validation, and test sets are constructed to be non-overlapping to prevent any data leakage. The detailed data generation process can be found in Appendix~\ref{apx:dynamics}.

\subsection{Baselines}
We compare our proposed FNFM against a comprehensive set of baselines, including standard forecasting models, adaptive methods, and an oracle-like expert model.
\begin{itemize}
    \item \textbf{STGCN}~\cite{yuspatiotemporal2018}: A deep learning model that combines graph convolutions for spatial dependencies with gated temporal convolutions for temporal dynamics, jointly modeling graph-structured time series.
    \item \textbf{STEP}~\cite{shaopretraining2022}: a novel framework that enhances spatialtemporal graph neural networks by using a pretraining model to learn temporal patterns and generate segment-level representations.
    \item \textbf{STGFSL}~\cite{luspatiotemporal2022}: A model-agnostic, spatiotemporal graph few-shot learning framework designed for scenarios.
    \item \textbf{GPD}~\cite{yuanspatiotemporal2024}: A generative pretraining framework that pretrains a diffusion model to generate customized parameters for spatiotemporal prediction networks.
    \item \textbf{One-per-Env STGCN}~\cite{yuspatiotemporal2018}: A set of expert STGCN models trained specifically for each environment, serving as a reference for the upper bound of performance.
\end{itemize}

\subsection{Main Results}
Table~\ref{tab:average_rmse} presents the comparative results on four datasets. 
Our proposed method, FNFM, consistently achieves state-of-the-art performance, outperforming all baselines across nearly every in-domain and out-of-domain scenario. 
Notably, FNFM's accuracy is highly competitive with the One-per-Env STGCN, an oracle-like expert model trained with full access to data from the target environment.

FNFM's effectiveness is attributable to its ability to explicitly learn the low-dimensional manifold of the expert model weights. 
While monolithic baselines seek a single compromise model and meta-learning requires target data to navigate this space, FNFM learns the global structure of the manifold itself. 
By using a VAE to identify this structure and a CFM to learn the direct map from any environmental coefficient to a point upon it, our framework can instantly generate a specialized, near-optimal model.

\begin{figure}[!t]
\centering
\subfigure[Epidemic]{
\includegraphics[width=0.207\textwidth]{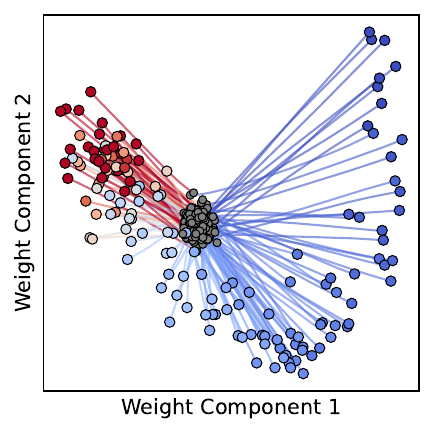}}
% \hspace{0.01\textwidth}
\subfigure[Collab]{
\includegraphics[width=0.254\textwidth]{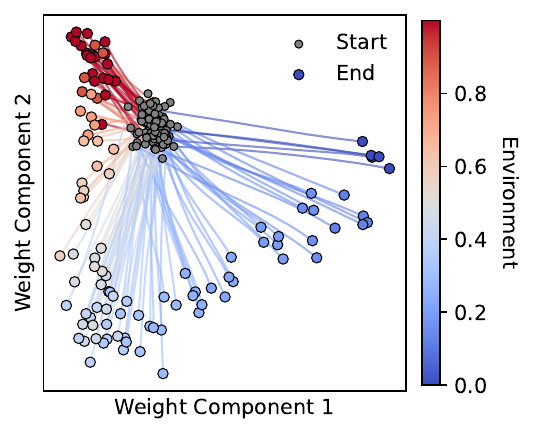}}
\vspace{-0.4cm}
\caption{The process of FNFM generating weights for a predictive model under different environmental conditions.}
\label{fig:generation}
\end{figure}

\subsection{Explainability}
To gain deeper insight into the internal workings of FNFM, we visualize its weight generation process on two datasets in Figure~\ref{fig:generation}. We use principal component analysis to project the high-dimensional latent space learned by the VAE onto a two-dimensional plane. The visualization clearly reveals three key elements of our framework's success.

First, the latent vectors of the expert models (the End points) are not scattered randomly; instead, they converge to form a smooth, well-structured, low-dimensional manifold. This indicates that the space of effective model weights possesses a strong intrinsic structure, which our VAE successfully captures.

Second, this manifold is meaningfully organized by the environmental coefficients, as illustrated by the color gradient. We observe a continuous color transition along the manifold's structure, signifying that similar environments correspond to proximate locations in the latent space. This confirms that our framework has learned a semantic mapping from dynamic environments to model weights.

Finally, the trajectories connecting the start points (from the Gaussian prior) to the end points (the target weights) visualize the conditional flow matching process. These trajectories follow direct, nearly-straight paths from the simple prior to their target locations on the manifold. This reveals that our CFM model has learned a stable and efficient transport map, ensuring high-quality zero-shot generation.

In summary, this visualization provides compelling evidence for FNFM's success: it learns not only a semantically organized manifold of expert models but also an effective conditional path to navigate it.

\begin{figure*}[!t]
    \centering
    \includegraphics[width=0.92\textwidth]{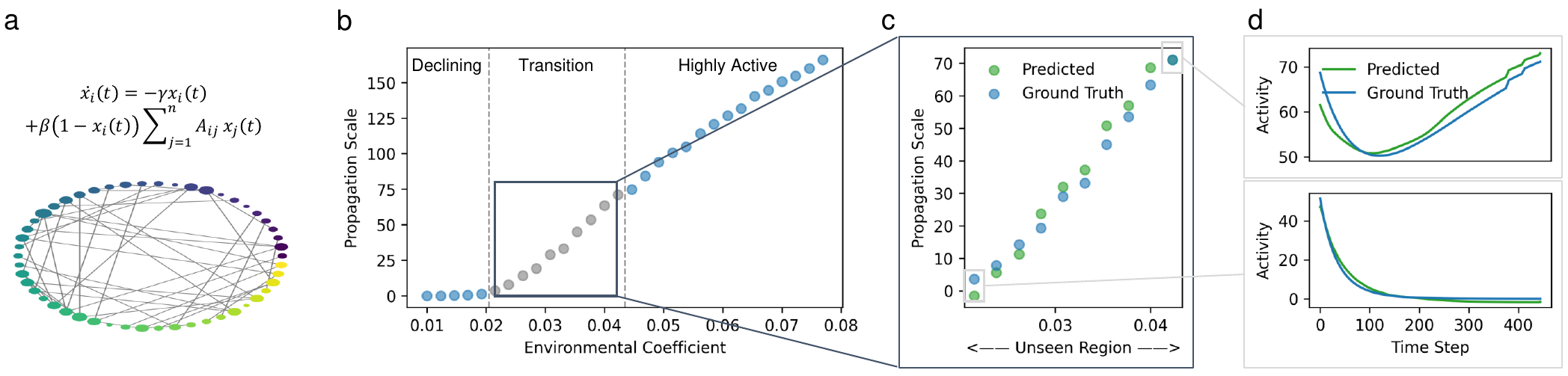}
    \caption{Case study on the Collab dataset. (a) Illustration of the network topology and governing equation. (b) The information propagation scale (network activity at the final time step) as a function of the environmental coefficient (popularity). (c) FNFM's generalized prediction of the propagation scale within the phase transition region closely matches the ground truth. (d) FNFM's predicted trajectories for two extreme scenarios: a declining case and an active case.}
    \label{fig:case_study}
\end{figure*}

\subsection{Case Study}
To further investigate FNFM's ability to generalize, we conduct a case study on the Collab dataset, which models an SIS-like information propagation process on a network of scientific collaborations (Figure~\ref{fig:case_study}a). This system exhibits a critical phenomenon known as a phase transition, where the long-term outcome of the dynamics is acutely sensitive to the environmental coefficient, which in this context represents the information's stickiness~\cite{vespignani2012modelling}.

As illustrated in Figure~\ref{fig:case_study}b, the system's final propagation scale displays three distinct regimes based on the environmental coefficient. When the coefficient is below a critical threshold (the Declining region), activity eventually dies out. Conversely, above a higher threshold (the Active region), the information becomes endemic, reaching a high, stable level of activity. Between these extremes lies a highly non-linear Transition region, where small changes in the coefficient lead to dramatic shifts in the outcome.

For this experiment, we deliberately trained FNFM only on data from the two extreme regimes (Declining and Active), leaving the entire critical Transition region as a challenging, unseen test bed. The results demonstrate a remarkable generalization capability. Figure~\ref{fig:case_study}c shows that FNFM's zero-shot predictions for the propagation scale in this unseen region align closely with the ground truth, indicating that our model successfully learned the underlying non-linear function governing the phase transition. Furthermore, Figure~\ref{fig:case_study}d displays two example trajectory forecasts, confirming that our generated models can accurately predict the full temporal evolution of the system's network-wide activity.

This case study provides strong evidence that FNFM learns more than simple input-output mappings; it captures the fundamental principles of a complex system's critical behavior. The ability to accurately interpolate within a phase transition showcases its potential as a powerful tool for reliable forecasts for systems with novel parameters near critical tipping points.

\begin{figure}[!t]
\centering
\subfigure[Data Ratio]{
\includegraphics[width=0.28\textwidth]{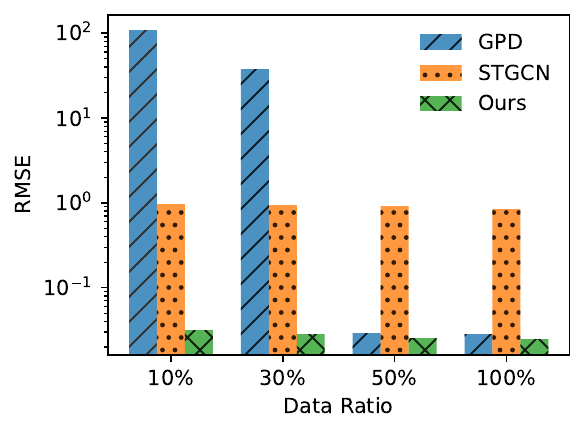}}
% \hspace{0.01\textwidth}
\subfigure[Noise]{
\includegraphics[width=0.18\textwidth]{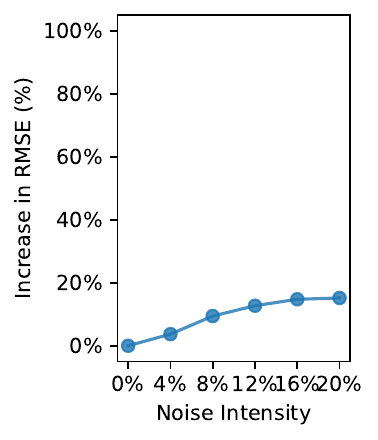}}
\vspace{-0.4cm}
\caption{Robustness results on the Collab dataset.}
\label{fig:robustness}
\end{figure}

\subsection{Robustness}
We conduct two additional experiments on the Collab dataset to assess the robustness of FNFM under challenging conditions: limited data availability and noisy environmental coefficients.

\paragraph{Robustness to Limited Data.}
In this experiment, we evaluate model performance when trained on a reduced number of available environments (from 100\% down to 10\%). As shown in Figure~\ref{fig:robustness}a, FNFM maintains its superior performance and low RMSE even when trained with only 10\% of the environments. In contrast, the competing generative model, GPD, suffers a catastrophic performance degradation under data scarcity, indicating its heavy reliance on a large number of training examples. This highlights FNFM's excellent data efficiency, suggesting that our framework can effectively learn the underlying manifold of expert weights from a very limited sample of environments.

\paragraph{Robustness to Noisy Coefficients.}
This experiment tests the model's stability when the provided environmental coefficients at inference time are inaccurate. We add zero-mean Gaussian noise (standard deviation is setting to the coefficient's total range) to the true coefficients, with the noise intensity varying from 0\% to 20\%. Figure~\ref{fig:robustness}b shows that the RMSE of FNFM increases gracefully and smoothly as the noise intensity grows, with less than a 16\% increase in error even at 20\% noise. This smooth degradation, rather than a sudden breakdown, provides strong evidence that FNFM has learned a continuous and well-behaved mapping from the coefficient space to the latent space of model weights. This property is crucial for practical applications, as it ensures that small estimation errors in the environmental coefficients will only lead to small and predictable errors in the final forecast.

\subsection{Ablation Study}
To validate the effectiveness of our key design choices, we conduct an ablation study on the Hill and Collab datasets, with results shown in Table~\ref{tab:ablation}.

\paragraph{Weight Sequence Tokenizer.}
This experiment assesses the importance of our structure-preserving tokenizer. In the ``w/o Tokenizer'' variant, we simply flatten all model weights into a single vector and then reshape it into a token sequence, disrupting the network's architectural inductive biases. The results show that this naive approach leads to a notable performance degradation on the Collab dataset, particularly in the out-of-domain split. This validates our hypothesis that preserving the computational structure of the weights is crucial for the generative model to learn a meaningful and generalizable representation.

\paragraph{Environmental Coefficients.}
This experiment evaluates the necessity of the conditional generation mechanism. In the ``w/o Condition'' variant, the CFM model is trained unconditionally to generate an "average" expert model. The results show a severe drop in performance across all datasets and splits. This confirms that the environmental coefficient is the essential guiding signal for synthesizing the correct, specialized model weights. Without this conditioning, the model fails to adapt to the specific dynamics of any given environment, underscoring the critical role of our conditional framework.

\begin{table}[!t]
\caption{Ablation studies on Hill and Collab datasets. `w/o' means `without'.}
\centering
\vspace{-0.3cm}
\resizebox{\columnwidth}{!}{%
\begin{tabular}{lcccc}
\midrule
& \multicolumn{2}{c}{Hill} & \multicolumn{2}{c}{Collab} \\
& in-domain & out-domain & in-domain & out-domain\\
\midrule
w/o Tokenizer & 17.3312  & 11.1209 & 0.0513 & 0.0359 \\
w/o Condition & 27.6424 & 17.5220 & 0.0670 & 0.0306 \\
Ours & 13.8942 & 8.5595 & 0.0475 & 0.0244 \\
\midrule
\end{tabular}%
}
\label{tab:ablation}
\end{table}

\section{Related work}
\subsection{Modeling of Network Dynamics}
Data-driven modeling of complex network dynamics, particularly with Graph Neural Networks (GNNs), has become a prominent research direction. Foundational models like STGCN \cite{yuspatiotemporal2018} and Graph WaveNet \cite{wugraph2019} established effective frameworks for spatio-temporal forecasting by integrating graph convolutions with temporal modeling. Subsequent research has advanced this field by incorporating more sophisticated mechanisms, such as using Neural Ordinary Differential Equations (ODEs) to capture continuous-time dynamics \cite{zang2020neural}, or designing specific encoders to handle dynamic topologies and generalize across different environments \cite{huang2021coupled, huang2023generalizing}.

However, a fundamental challenge for these models is generalizing to unseen dynamic regimes. To address this, one line of work focuses on building large-scale, "one-for-all" foundation models that train on diverse data \cite{shaopretraining2022,li2024predicting}. While powerful, they often sacrifice the specialized accuracy required for specific dynamic conditions. Another prominent approach leverages meta-learning to quickly adapt a base model to new environments \cite{finnmodelagnostic2017, luspatiotemporal2022}. Despite their adaptability, these methods typically require at least a small amount of trajectory data from the target environment for finetuning, precluding their use in true zero-shot scenarios. Our work targets this critical gap.

\subsection{Generative Models for Network Weights}
Generating neural network weights is an emerging paradigm with significant potential for generalization \cite{wang2024neural}. This area has evolved along several fronts. One initial line of work focused on generating weights to accelerate or improve the training process itself, effectively replacing hand-crafted initializations \cite{gong2024efficient, schurholt2022hyper}.

More recent studies, closer to our own, leverage conditional generative models like diffusion to produce weights tailored for generalization. For instance, \citet{yuanspatiotemporal2024} use an urban knowledge graph as a prompt to generate spatio-temporal models for unseen cities. Others have integrated weight generation into the meta-learning loop, replacing gradient-based inner-loop updates with a diffusion process \cite{zhang2024metadiff}. While these methods represent significant progress, their zero-shot performance is often limited, as many still necessitate post-generation finetuning.

A potential reason for this limitation lies in the representation of the weights themselves. Most existing methods treat the weights as a simple flat vector, disrupting the network's inherent architectural inductive biases and making the distribution harder for a generative model to learn. In contrast, our work introduces two key innovations: 1) We employ a novel weight sequence tokenizer that preserves the computational structure of the network, providing a more meaningful representation for the generative model. 2) By synergistically combining a VAE and conditional flow matching, we directly learn a smooth manifold of expert weights, enabling truly zero-shot, single-pass generation of high-performance models without any subsequent tuning.

\section{Conclusion}
This work introduced FNFM, a novel generative framework that successfully addresses the critical challenge of zero-shot prediction for network dynamics. Recognizing that dynamics are highly sensitive to their governing coefficients, we proposed a paradigm shift: from directly predicting trajectories to generating the complete weights of a specialized forecaster model itself. Our methodology operationalizes this concept through a synergistic combination of a variational autoencoder and a conditional flow matching model. The VAE first learns a compact and smooth latent manifold of expert model weights, after which the CFM framework learns to map any environmental coefficient to a specific location on this manifold. This mechanism enables the instantaneous, single-pass generation of tailored models for unseen environments, demonstrating robust generalization without any need for finetuning.
\paragraph{Limitations \& Future work.}
A limitation is that our current framework assumes a static network topology; extending FNFM to handle dynamic graphs where nodes and edges evolve over time is a critical next step.

\bibliographystyle{ACM-Reference-Format}
\bibliography{reference}

\appendix

\section{Datasets}

subsection{Network Dynamics} \label{apx:dynamics}

\textbf{Hill dynamics} \cite{hensspatiotemporal2019} describes regulatory interactions in sub-cellular networks, with the following governing equations
\begin{equation}\label{Hill}
    \frac{dx_{i}}{dt}=-B_{i}x_{i}^{a}+\sum_{j=1}^{N}A_{ij}\frac{x_{j}^{h}}{1+x_{j}^{h}} ,
\end{equation}
where $x_i(t)$ is the abundance of protein $i$, $A_{ij}$ is the network topology, and the exponents $a$ and $h$ control the self-dynamics and regulatory interaction, respectively.

\textbf{SIS dynamics} \cite{allen1994some,vespignani2012modelling} models an epidemic process on a network where nodes can be infected by their neighbors and recover to a susceptible state. The governing equation for the infection probability $x_i$ of each node $i$ is
\begin{equation}\label{SIS}
    \frac{dx_i}{dt} = -\gamma x_i + \beta (1-x_i) \sum_{j=1}^{N} A_{ij}x_j ,
\end{equation}
where $x_i$ is the state of node $i$, $A_{ij}$ is the adjacency matrix representing the network topology, $\gamma$ is the recovery rate, and $\beta$ is the infection rate per contact.

\subsection{Simulation}
We employ Euler's method to numerically solve the dynamics of the aforementioned systems for each topology, generating evolutionary trajectories as the datasets. To create challenging generalization tasks, we vary the key dynamic coefficients to form distinct training and testing environments. 
For the Hill dataset, the out-of-domain (OOD) environments are created by sampling coefficients from the boundaries of the training ranges; specifically, setting parameter $a=0.6$ (from a training range of $[0.5, 0.6]$) and sampling parameter $h$ from the sub-range $[1.2580, 2.0000]$ (from a full training range of $[0.33, 2.00]$).
For the Epidemic and Twitter datasets, the parameter $\beta$ is varied across its full range while the coefficient $\gamma$ is split. 
For Epidemic, the training range for $\gamma$ is $[0.0200, 0.0330]$, while the OOD test set extrapolates to $[0.0360, 0.0390]$.
For Twitter, the training range for $\gamma$ is $[0.0200, 0.0740]$, with its OOD test set in the extrapolated range of $[0.0740, 0.0800]$.
Finally, for the Collab dataset, the parameter $\beta$ is fixed at $0.02$, while the training environments use a $\gamma$ range of $[0.2000, 0.4264]$. The OOD test set for this dataset involves a distributional shift to a completely separate, non-overlapping range of $[0.4728, 0.9302]$.

\section{Software and Hardware Environment}

We implement FNFM in PyTorch and employ the open-source available implementations with default parameters for baselines. All experiments were conducted on the NVIDIA GeForce RTX 2080Ti GPU. For all datasets and baselines, we set the batch size to 64 and trained for 500 epochs with a learning rate of 0.0001.

\section{Model Configuration}
\paragraph{Weight Variational Autoencoder.}
Our Weight Variational Autoencoder (VAE) is built upon a Transformer architecture designed to process the sequence of model weights. The VAE's internal model dimension ($d_{model}$) is set to 128. The Transformer architecture consists of 2 layers, each equipped with 8 attention heads. The encoder network compresses the input weight token sequence into a 32-dimensional latent space. For training, we used the Adam optimizer with an initial learning rate of 1e-4 and a weight decay of 3e-9, managed by a OneCycleLR scheduler. The batch size was set to 32, and the Kullback-Leibler (KL) divergence term in the ELBO loss was weighted by a $\beta$ factor of 1e-6.

\paragraph{Conditional Flow Matching Model.}
The conditional vector field ($v_\xi$) for the Flow Matching process is also implemented using a Transformer-based architecture. This network operates on the 32-dimensional latent space, taking a latent vector as input and outputting a velocity vector of the same dimension. The architecture is composed of 4 Transformer layers, each with 2 attention heads. A dropout rate of 0.1 is applied during training for regularization. The model is conditioned on external environmental information, which is provided through dedicated knowledge graph and time embeddings.

\end{document}